\documentclass[prl,twocolumn,preprintnumbers,amsmath,amssymb,amsfonts]{revtex4}
\usepackage{color}
\usepackage{refstyle}
\usepackage{mathrsfs}
\usepackage{esint}
\usepackage{bm}
\usepackage{cancel}
\usepackage[normalem]{ulem}
\usepackage{graphicx}
\usepackage[caption=false]{subfig}
\usepackage[unicode=true,pdfusetitle,bookmarks=true,
bookmarksnumbered=false,bookmarksopen=false,breaklinks=false,
pdfborder={0 0 0},backref=false,colorlinks=true,citecolor=red]
{hyperref}
\usepackage{cleveref}
\providecommand\eqref[1]{\ref{eq:#1}}
\renewcommand\b[1]{{\bf  #1}}
\renewcommand\vec[1]{\boldsymbol{#1}}
\renewcommand\phi{\varphi}
\newcommand\tr{\mathrm{tr}}
\newcommand\del{\nabla}
\newcommand\dd{\mathrm{d}}

\allowdisplaybreaks[3]

\begin{document}
\title{Defect Unbinding in Active Nematics}
\author{Suraj Shankar$^{a,b}$}
\email[]{sushanka@syr.edu}
\author{Sriram Ramaswamy$^{b,c}$}
\email[]{sriram@iisc.ac.in}
\author{M. Cristina Marchetti$^{a,b}$}
\email[]{mcmarche@syr.edu}
\author{Mark J. Bowick$^{a,b}$}
\email[]{bowick@kitp.ucsb.edu}
\affiliation{$^a$Physics Department and Syracuse Soft and Living Matter Program, Syracuse University, Syracuse, NY 13244, USA.\\
$^b$Kavli Institute for Theoretical Physics, University of California, Santa Barbara, CA 93106, USA.\\
$^c$Centre for Condensed Matter Theory, Department of Physics, Indian Institute of Science, Bangalore 560 012, India.}
\date{\today}
\begin{abstract}
We formulate the statistical dynamics of topological defects in the active nematic phase, formed in two dimensions by a collection of self-driven particles on a substrate. An important consequence of the non-equilibrium drive is the spontaneous motility of strength $+1/2$ disclinations.
	Starting from the hydrodynamic equations of active nematics, we derive an interacting particle description of defects that includes active torques. We show that activity, within perturbation theory, lowers the defect-unbinding transition temperature, determining a critical line in the temperature-activity plane that separates the quasi-long-range ordered (nematic) and disordered (isotropic) phases. Below a critical activity, defects remain bound as rotational noise decorrelates the directed dynamics of $+1/2$ defects, stabilizing the quasi-long-range ordered nematic state.
This activity threshold vanishes at low temperature, leading to a re-entrant transition. At large enough activity, active forces always exceed thermal ones and the perturbative result fails, suggesting that in this regime activity will always disorder the system. Crucially, rotational diffusion being a two-dimensional phenomenon, defect unbinding cannot be described by a simplified one-dimensional model.
%
\end{abstract}
\maketitle
	Liquid crystals exhibit remarkable orientationally ordered phases, the simplest being the nematic phase in which particles macroscopically align along a single preferred orientation, without a head-tail distinction. The name nematic itself comes from $\nu\eta\mu\alpha$, meaning thread, for the line-like topological defects (disclinations) that are inevitably produced in quenches from the high-temperature disordered phase to the nematic phase \cite{de1995physics,chuang1991cosmology,bowick1994cosmological,volovik2003universe}. In two dimensions ($2d$), though, disclinations are point-like defects, and so may be thought of as localized particles. The nematic pattern around a disclination is a distinctive fingerprint of the spontaneous symmetry-breaking that characterizes nematic order and distinguishes the elementary defects from, say, integer strength vortices in two-dimensional spin systems. The nematic director rotates through a half-integer multiple of $2\pi$ as one circumnavigates a defect. Thus, the lowest-energy defects are referred to as carrying strength $\pm1/2$. In two dimensional equilibrium nematics the entropic unbinding of such point disclinations drives the nematic to isotropic (NI) transition \cite{berezinskii1971destruction,kosterlitz1973ordering,kosterlitz1974critical,stein1978kosterlitz}.
	
\begin{figure}[]
	\centering
	\includegraphics[width=0.45\textwidth]{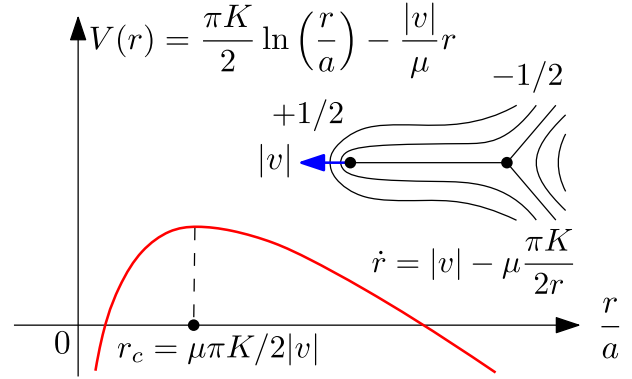}
	\caption{Potential $V(r)$ for a neutral defect pair for the configuration in which the direction of motility of the $+1/2$ disclination points away from the $-1/2$ and is held fixed. This na{\"i}ve picture suggests that incipient active defect pairs have an exponentially small, but finite, rate to overcome the barrier at low temperature, and hence always unbind.}
	\label{fig:strawman}
\end{figure}

In recent years there has been much focus on nematics composed of elongated units that are self-driven - hence \textit{active} nematics 
~\cite{ramaswamy2003active,ahmadi2006hydrodynamics}. Examples include collections of living cells \cite{gruler1999nematic,kemkemer2000elastic,duclos2014perfect,Zhou2014,nishiguchi2017long,kawaguchi2017topological,saw2017topological}, synthetic systems built of cellular extracts~\cite{sanchez2012spontaneous,keber2014topology,ellis2018curvature}, and vibrated granular rods~\cite{narayan2007long}. Active nematics exhibit complex spatio-temporal dynamics, accompanied by spontaneous defect proliferation. Much progress has been made in understanding the properties of the ordered phase \cite{ramaswamy2003active,mishra2010dynamic,bertin2013mesoscopic,shi2014instabilities,ngo2014large,shankar2018low}, but a complete theory of order, fluctuations, defects and phase transitions of active nematics still eludes us. Although the nematic itself has no net polarity, the director pattern around a strength $+1/2$ defect has a local comet-like geometric polarity (Fig.~\ref{fig:strawman}). In an active system this renders $+1/2$ defects motile~\cite{giomi2013defect,narayan2007long} with a self-propelling speed proportional to activity \cite{giomi2013defect}. Both experiments \cite{sanchez2012spontaneous,keber2014topology,Zhou2014,guillamat2016probing,guillamat2016control,guillamat2017taming,ellis2018curvature} and simulations \cite{giomi2013defect,thampi2013velocity,shi2013topological,giomi2014defect,thampi2014instabilities,thampi2014vorticity,gao2015multiscale,giomi2015geometry} have shown that motile defects play a key role in driving self-sustained active flows.

In this Letter we precisely map the dynamics of active defects onto that of a mixture of motile ($+1/2$) and passive ($-1/2$) particles with interaction forces and aligning torques, putting on firm ground previous purely phenomenological models~\cite{giomi2013defect,pismen2013dynamics,keber2014topology}.~A key new result is the derivation of the angular dynamics of the $+1/2$ defects. Treating activity as a small parameter, we then construct and solve the defect Fokker-Planck equation and show that activity weakens the logarithmic attraction between opposite-charge defects. As a result, increasing activity past a threshold drives a nonequilibrium NI phase transition to a phase of unbound defects, much like the Berezenskii-Kosterlitz-Thouless (BKT) transition in two-dimensional spin systems~\cite{berezinskii1971destruction,kosterlitz1973ordering,kosterlitz1974critical} and passive nematics~\cite{stein1978kosterlitz}.
Rotational diffusion ($D_R$) of the $+1/2$ defect is suppressed at low noise, where self-propulsion directly drives unbinding with a threshold that vanishes as $D_R$ goes to zero. This yields
a re-entrant isotropic-nematic-isotropic sequence \footnote{A. Maitra and M. Cates, personal communication.} as a function of temperature at fixed activity.
Our effective equations for defect dynamics also provide a simple model capable of quantifying the dynamics of interacting active defects in confined geometries.

	The proof of the existence of a low-activity quasi-long-range ordered active nematic in $2d$ \cite{mishra2010dynamic,shankar2018low} is an important result because a na{\"i}ve argument suggests otherwise.
	In an equilibrium nematic, two $\pm1/2$ defects at a distance $r$ experience an attractive interaction $V_0(r)= (\pi K/2)\ln\left(r/a\right)$, with $K$ a Frank elastic constant and $a$ the size of the defect core. Hence, neglecting inertia, they are drawn towards each other according to $\dot{r}=-\mu\partial_rV_0$, with $\mu$ a defect mobility. One could then argue that the dynamics of a suitably oriented $\pm 1/2$ defect pair in an \emph{active} nematic
is governed by relaxation in an effective potential \cite{giomi2013defect}
\begin{equation}
	\dot{r}=-\mu\partial_rV\ ,\quad V(r)=\dfrac{\pi K}{2}\ln\left(\dfrac{r}{a}\right)-\dfrac{|v|}{\mu}r\ ,
\end{equation}
where $|v|$ is the self-propelling speed with which the $+1/2$ disclination is moving away from the $-1/2$ disclination (see Fig.~\ref{fig:strawman}).
	The resulting barrier $V(r_c)=(\pi K/2)\left[\ln\left(\pi\mu K/(2|v|a)\right)-1\right]$ at distance $r_c=\pi\mu K/(2|v|)$ is finite, which means that the defect pair is always unbound, and active nematic order thus destroyed, at any nonzero temperature (Fig.~\ref{fig:strawman}). As activity is increased, more and more defect pairs will be liberated \cite{sanchez2012spontaneous,thampi2013velocity,giomi2013defect} suggesting that nematic order would be completely destroyed by the swarming disordered cores, much like driven vortices in superconducting films can destroy superconductivity.
	Here we show that this heuristic argument fails because rotational noise, by disrupting the directed motion of the $+1/2$ defects, counterintuitively restores the ordered nematic phase.

	We begin with the hydrodynamics of a two-dimensional nematic liquid crystal written in terms of the flow velocity $\b{u}$ and the tensor order parameter $Q_{\mu\nu}=S(2\hat{n}_\mu\hat{n}_\nu-\delta_{\mu\nu})$, where $S$ is the scalar order parameter and $\hat{\b{n}}$ is the director field. We ignore density fluctuations,
although we expect this restriction could be dropped without qualitatively changing the results. The $\b{Q}$ equation is as for passive nematics \cite{beris1994thermodynamics},
\begin{equation}
	\gamma\mathcal{D}_t\b{Q}=\left[a_2-a_4\ \tr(\b{Q}^2)\right]\b{Q}+K\del^2\b{Q}\ ,\label{eq:Qeqn}
\end{equation}
where $\mathcal{D}_t=\partial_t+\b{u}\cdot{\bm\del}-[\cdot,\b{\Omega}]$ is the comoving and corotational derivative with the vorticity tensor $\Omega_{\mu\nu}=(\nabla_{\mu}u_{\nu}-\nabla_{\nu}u_{\mu})/2$. Only the relaxational part of the dynamics is retained in Eq.~\ref{eq:Qeqn}, with $\gamma$ a rotational viscosity, $K$ a Frank elastic constant and $a_2,a_4$ the
parameters that set the mean-field NI transition at $a_2=0$. A treatment including various flow alignment terms is given in the SI. At equilibrium, the homogeneous ordered state for $a_2>0$ has $S_0=\sqrt{a_2/(2a_4)}$ and an elastic coherence length $\xi=\sqrt{K/a_2}$. For an isolated static $\pm1/2$ defect in equilibrium, the director $\hat{\b{n}}(\phi)=\left(\cos(\phi/2),\pm\sin(\phi/2)\right)$ rotates by $\pm\pi$ with the azimuthal angle $\phi$, and $S$ vanishes at the core of the defect, assuming its bulk value on length scales larger than the defect core size $a\sim\xi$.
Activity enters in the force balance equation, which, ignoring inertia and in-plane viscous dissipation, is given by $-\Gamma\b{u}+{\bm\del}\cdot\vec{\sigma}^a=\b{0}$, where $\Gamma$ is the friction with the substrate and $\vec{\sigma}^a=\alpha\b{Q}$ is the active stress tensor that captures the internal forces generated by active units~\cite{ramaswamy2010mechanics,marchetti2013hydrodynamics}. We neglect elastic and Ericksen stresses as they are higher order in gradients. The system is extensile for $\alpha<0$ and contractile for $\alpha>0$. For a $+1/2$ disclination, the active backflow at its core gives rise to a self-propulsion speed $\sim|\alpha|/(\Gamma a)$~\cite{giomi2013defect,pismen2013dynamics}.

	The $+1/2$ disclination has a local geometric polarization $\b{e}_i=a{\bm\del}\cdot\b{Q}(\b{r}_i^+)$ (evaluated at the core of the defect), defined here to be dimensionless. Note that $\b{e}_i$ is \emph{not} a unit vector. Our treatment does not require the mode expansion used in Ref.~\cite{cortese2018pair} to treat multi-defect configurations. An isolated $+1/2$ defect has a non-vanishing flow velocity at its core ($\b{u}(\b{r}_i^{+})=v\b{e}_i$, $v=\alpha S_0/\Gamma a$), while the $-1/2$ defect does not, due to its three-fold symmetry ($\b{u}(\b{r}^-_i)=\b{0}$) \footnote{Including a ``second'' active force $\sim\alpha'\b{Q}\cdot(\bm\del\cdot\b{Q})$ does not affect the ballistic motion of the $+1/2$ defect \cite{maitra2018nonequilibrium}.}. We show that the resulting positional dynamics of the defects, including both motility and passive interactions (for a derivation, see SI) \footnote{Apart from the motility of the $+1/2$ defect, both charge disclinations are also entrained by active flows generated by other defects. This leads to $\dot{\b{r}}^{\pm}_i\sim\alpha\vec{\sigma}_3\cdot{\bm\del}_i\sum_{j\neq i}q_j\ln|(\b{r}_i-\b{r}_j)/a|$, where $\vec{\sigma}_3$ is a Pauli matrix, and provides an anisotropic active correction to bend and splay elasticity. Including fluctuations, this term is $\sim\mathcal{O}(r_{ij}^{-1-\eta})$ \cite{shankar2018low} where $r_{ij}$ is the distance between two defects and $\eta=T/(2\pi K)$, so it is subdominant to the passive interaction ${\bm\del}_i\mathcal{U}$.}, is given by
\begin{subequations}
\begin{gather}
	\dot{\b{r}}^+_i=v\b{e}_i-\mu{\bm\del}_{i}\mathcal{U}+\sqrt{2\mu T}\vec{\xi}_i(t)\ ,\label{eq:+1/2eom}\\
	\dot{\b{r}}^-_i=-\mu{\bm\del}_{i}\mathcal{U}+\sqrt{2\mu T}\vec{\xi}_i(t)\ ,\label{eq:-1/2eom}
\end{gather}\label{eq:eom}
\end{subequations}
\noindent where $\mu\propto1/\gamma$ is a defect mobility, $\vec{\xi}_i(t)$ Gaussian white noise and
\begin{equation}
	\mathcal{U}=-2\pi K\sum_{i\neq j}q_iq_j\ln\left|\dfrac{\b{r}_i-\b{r}_j}{a}\right|\ ,\label{eq:logpot}
\end{equation} 
is the Coulomb interaction between defects, with $q_i=\pm1/2$ the strength of the $i$th defect. The elastic constant $K$ includes corrections from hydrodynamic flows linear in activity which can destabilize the nematic state even in the absence of topological defects~\cite{srivastava2016negative,maitra2018nonequilibrium}. Here we take $K>0$ (permitted in a domain of parameter space~\cite{srivastava2016negative,maitra2018nonequilibrium}) to guarantee an elastically stable nematic.
Note that $v\propto\alpha$ can be of either sign. The translational noise strength $T$ arises from thermal or active noise in the $\b{Q}$ equation (Eq.~\ref{eq:Qeqn}). A more sophisticated calculation (see SI)
gives logarithmic corrections to the defect mobility $\mu$~\cite{pismen1990mobility,ryskin1991drag,denniston1996disclination,pismen1999vortices}. The important feature of activity is that it elevates the geometric structural polarity of the $+1/2$ disclination to a \emph{dynamical} degree of freedom, one that drives motion. In turn, $\b{e}_i$ also has its own dynamics, which is, in principle, contained in the $\b{Q}$ equation. Neglecting noise for now and using the quasistatic approximation in a frame comoving with the $+1/2$ defect, i.e $[\partial_t\b{Q}]_{\b{r}_i^+(t)}=\b{0}$, we have $\dot{\b{e}}_i(t)=a[\b{v}_i(t)\cdot{\bm\del}]{\bm\del}\cdot\b{Q}(\b{r}_i^+(t))$, where $\b{v}_i=v\b{e}_i-\mu{\bm\del}_i\mathcal{U}$ is the deterministic part of $\dot{\b{r}}_i^+$ (Eq.~\ref{eq:+1/2eom}). Our approximation neglects elastic torques on $\b{e}_i$ due to smooth director distortions, shown to be unimportant for the dynamics of neutral pairs \cite{vromans2016orientational,tang2017orientation} (a more detailed justification and comparison with previous work is given in the SI).
Assuming a dilute gas of slowly moving defects, we perturbatively expand Eq.~\ref{eq:Qeqn} about the equilibrium defect configuration and solve for $\b{Q}$. Using this solution, we evaluate ${\bm\del}{\bm\del}\cdot\b{Q}$ at the core of the defect to obtain (for details, see SI)
\begin{equation}
	\dot{\b{e}}_i=-\dfrac{5\gamma}{8K}\left[\b{v}_i\cdot\left(\b{v}_i-v\b{e}_i\right)\right]\b{e}_i-\dfrac{v\gamma}{8K}(\b{v}_i\times\b{e}_i)\ \vec{\epsilon}\cdot\b{e}_i\ ,
	\label{eq:edot-det}
\end{equation}
where $\vec{\epsilon}$ is the two-dimensional Levi-Civita tensor.
Since $\b{e}_i$ is not a unit vector, its deterministic dynamics has a term along $\b{e}_i$ fixing its preferred magnitude and one transverse to it aligning the polarization to the force. 

\begin{figure}[]
	\centering
	\includegraphics[width=0.5\textwidth]{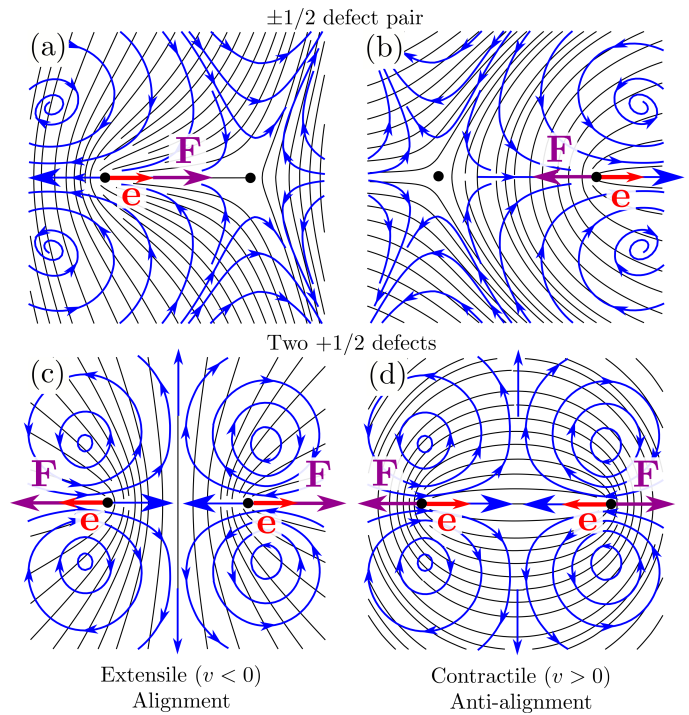}
	\caption{Configurations of defect pairs whose orientations, for an imposed fixed separation, are stable to transverse fluctuations of the $+1/2$ polarization(s). The active backflow is shown in blue and the director configuration in black. The polarization and force on each $+1/2$ defect is shown in red and in purple respectively. The top row shows a neutral $\pm1/2$ defect pair orientationally stable for (a) extensile ($v<0$) and (b) contractile ($v>0$) systems. Similarly, in the bottom row we have a pair of $+1/2$ defects that are orientationally stable. The far field nematic texture for these two-defect configurations has an aster-like structure when (c) extensile ($v<0$) and a vortex-like structure when (d) contractile ($v>0$).}
	\label{fig:torque}
\end{figure}

To elucidate the nature of the torques on the polarization, we write $\b{e}_i=|\b{e}_i|(\cos\theta_i,\sin\theta_i)$ and decompose the elastic force acting on the $i^{\mathrm{th}}$ defect ($\b{F}_i=-{\bm\del}_i\mathcal{U}$) as $\b{F}_i=|\b{F}_i|(\cos\psi_i,\sin\psi_i)$. For the defect orientation $\theta_i$, Eq.~\ref{eq:edot-det} then reduces to
\begin{equation}
	\partial_t\theta_i=v\dfrac{\mu\gamma}{8K}|\b{F}_i||\b{e}_i|\sin(\theta_i-\psi_i)\ .\label{eq:torque}
\end{equation}
 Active backflows tend to align the defect polarization with the force acting on the defect. A similar alignment kernel has been used previously to phenomenologically model flocking and jamming in cellular systems \cite{szabo2006phase,henkes2011active}, but here it arises naturally from the active dynamics of a two-dimensional nematic. Importantly, here the torque is controlled by activity ($v\propto\alpha$). An extensile system ($v\propto\alpha<0$) favors \emph{alignment} of the polarization with the force, while a contractile system ($v\propto\alpha>0$) favors \emph{anti-alignment} of polarization and force (Fig.~\ref{fig:torque}). The equations obtained here also predict patterns for four $+1/2$ defects on a sphere as obtained in Ref.~\cite{keber2014topology}.

For configurations in which the $+1/2$ is running \emph{away} from the $-1/2$ in an isolated neutral defect pair, the active aligning torque (Eq.~\ref{eq:torque}) stabilizes the $+1/2$ defect polarization against transverse fluctuations (see Fig.~\ref{fig:torque}a-b), irrespective of the sign of activity. Hence activity not only renders the $+1/2$ defect motile, but enhances the persistence of defect motion through the torques, favoring the unbinding of defect pairs.
This feature breaks the symmetry between pair creation and annihilation events for both extensile and contractile systems and justifies the one-dimensional cartoon in Fig.~\ref{fig:strawman}. As we will see below, however, the stochastic part of the defect dynamics (neglected so far) can disrupt these configurations, preventing unbinding. We finally remark that one can also obtain configurations for pairs of $+1/2$ disclinations (Fig.~\ref{fig:torque}c-d) that are stable against transverse deflections of either polarization. As shown, in the far-field of these two-defect configurations, aster-like structures are favored in an extensile system while vortex-like structures are favored in a contractile one, as seen in confined fibroblasts~\cite{duclos2017topological}.

The stochastic part of the dynamics of $\b{e}_i$ also derives from noise in the dynamics of $\b{Q}$, but a full calculation is challenging and beyond the scope of the present work. In the limit of low activity, we assume that the noise statistics can be inferred from the known equilibrium joint probability distribution of ${\b{r}}_i^{\pm}$ and $\b{e}_i$,
\begin{equation}
P_{\mathrm{eq}}^{2N}=\dfrac{1}{Z_{2N}}e^{-\mathcal{U}/T}\prod_{i=1}^N\left(\dfrac{ K}{2\pi T}e^{-K|\b{e}_i|^2/2T}\right)\ , \label{eq:defectBoltz}
\end{equation}
where $Z_{2N}$ is the Coulomb gas partition function and $K|\b{e}_i|^2/2$ is the simplest contribution to the defect core energy \cite{chaikin2000principles}. This results in
\begin{align}
	\dot{\b{e}}_i&=\dfrac{5\mu\gamma}{8K}\left[{\bm\del}_i\mathcal{U}\cdot\left(v\b{e}_i-\mu{\bm\del}_i\mathcal{U}\right)\right]\b{e}_i+\dfrac{v\mu\gamma}{8K}({\bm\del}_i\mathcal{U}\times\b{e}_i)\ \vec{\epsilon}\cdot\b{e}_i\nonumber\\
	&\quad-\sqrt{2D_R}\ \vec{\epsilon}\cdot\b{e}_i\ \eta_i(t)+\vec{\nu}_i(t)\ ,\label{eq:edot}
\end{align}
where we have written $\b{v}_i$ in terms of the force $-{\bm\del}_i\mathcal{U}$.
Smooth director phase fluctuations can be shown to generate rotational noise (first term in the second line of Eq.~\ref{eq:edot}) that changes the direction of $\b{e}_i$, while keeping $|\b{e}_i|$ fixed. Here $\eta_i(t)$ is unit white noise and $D_R=\mu T/\ell_R^2$ is the rotational diffusivity of the $+1/2$ defect, with $\ell_R\sim a$. The properties of the longitudinal component $\vec{\nu}_i(t)$ of the noise are determined by requiring 
%
%
that the probability distribution of the defect gas relaxes to the corresponding equilibrium form where (for one Frank constant) defect position and polarization are decoupled
in the absence of activity (i.e., for $v=0$), with the result (see SI)
\begin{equation}
\langle\vec{\nu}_i(t)\vec{\nu}_j(t')\rangle=\b{1}\delta_{ij}T\dfrac{5\mu^2\gamma}{4}\dfrac{|{\bm\del}_i\mathcal{U}|^2}{K^2}\delta(t-t')\ .\label{eq:nunoise}
\end{equation}
No summation on repeated indices is implied. As written, the noise has no stochastic ambiguity and is independent of any thermodynamic parameters, involving only the defect mobility $\mu$ and rotational viscosity $\gamma$, as it should.

\begin{figure}[]
	\centering
	\includegraphics[width=0.45\textwidth]{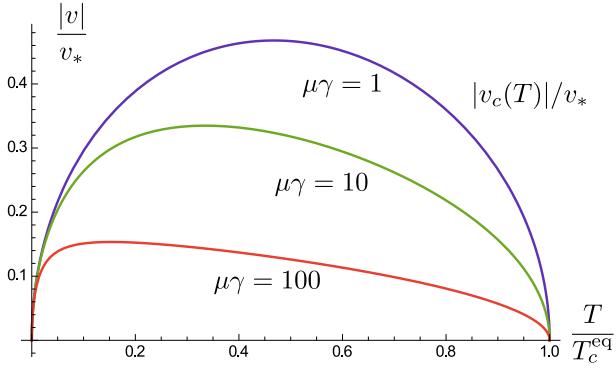}
	\caption{Phase boundary in the $|v|-T$ plane (Eq.~\ref{eq:v0c}) for different values of $\mu\gamma$. The region enclosed by the curve $|v_c(T)|$ for a given $\mu\gamma$ corresponds to the ordered nematic.}
	\label{fig:phase_dgm}
\end{figure}

To study defect unbinding, we now examine the dynamics of an isolated $\pm1/2$ defect pair governed by coupled Langevin equations for the pair separation $\b{r}=\b{r}^+-\b{r}^{-}$ (obtained from Eqs.~\ref{eq:+1/2eom},\ref{eq:-1/2eom}) and the $+1/2$ polarization $\b{e}$ (Eq.~\ref{eq:edot}). We derive and solve the corresponding Fokker-Planck equation for the steady state distribution, perturbatively in activity by using an isotropic closure for $\langle\b{e}\b{e}\rangle$ and neglecting all higher order moments in $\b{e}$ (see SI). Integrating over the polarization, we obtain the steady-state defect pair density at large distances to have an equilibrium-like form $\rho_{ss}(\b{r})\propto e^{-\mathcal{U}_{\mathrm{eff}}(\b{r})/T}$ with an effective
pair potential $\mathcal{U}_{\mathrm{eff}}(\b{r})\simeq (\pi K_{\mathrm{eff}}/2)\ln(r/a)$ where, to leading order in activity,
\begin{equation}
	K_{\mathrm{eff}}(v)=K-\dfrac{v^2}{2\mu D_R}\left[1+\mu\gamma\dfrac{3T}{4K}\right]+\mathcal{O}(v^4)\ .
\label{eq:Keff}
\end{equation}

\begin{figure}[]
	\centering
	\includegraphics[width=0.5\textwidth]{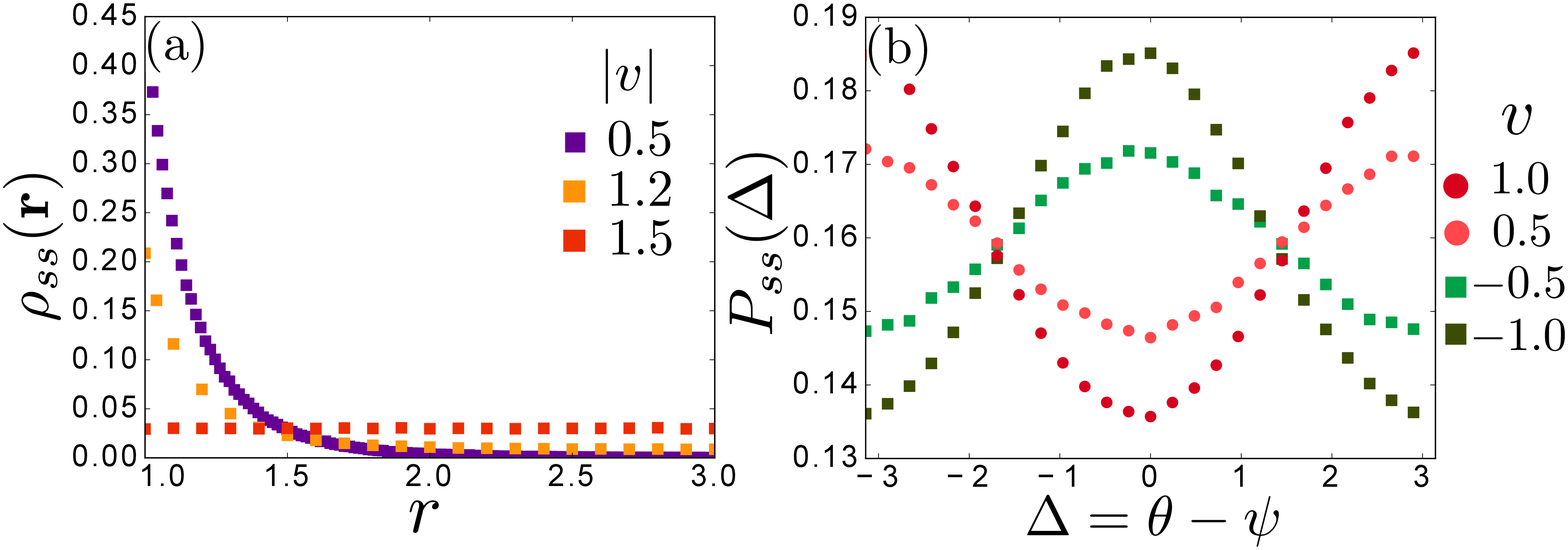}
	\caption{Steady-state statistics for a $\pm1/2$ defect pair in a periodic box of size $L=50 a$ ($T/T_c^{\mathrm{eq}}=0.51$, all other parameters are unity). (a) The pair separation distribution $\rho_{ss}(\b{r})$ for low ($|v|=0.5,1.2$, bound phase) and high ($|v|=1.5$, unbound phase) activity, suggesting that Eq.~\ref{eq:v0c} which gives $|v_c|\simeq2.06$, overestimates the unbinding threshold. (b) The distribution of the relative angle ($\Delta=\theta-\psi$) between the polarization $\b{e}$ and the force $\b{F}$ on the $+1/2$ defect for extensile ($\square$) and contractile ($\bigcirc$) systems.}
	\label{fig:dist}
\end{figure}
Hence, for large pair separation, the defect interaction is weakened by activity.
A small activity reduces the entropic BKT transition temperature $T_c^{\mathrm{eq}}=\pi K/8$ to $T_{c}(v)=\pi K_{\mathrm{eff}}(v)/8$. Inverting this equation for small $|v|$, we obtain the phase boundary below which the ordered nematic is stable,
\begin{equation}
	\dfrac{|v_c(T)|}{v_*}=\sqrt{\dfrac{16\ \tilde T(1-\tilde T)}{\pi\left[1+(3\pi/32)\mu\gamma\tilde T\right]}}\ ,\label{eq:v0c}
\end{equation}
with $\tilde T=T/T_c^{\mathrm{eq}}$ and $v_*=\mu T_c^{\mathrm{eq}}/\ell_R$. As shown in Fig.~\ref{fig:phase_dgm}, this implies re-entrant behavior as a function of $T$.
If the rotational diffusivity $D_R$ has a non-thermal part $D_R^a$, then there is a nonzero activity threshold $\sim\sqrt{D_R^a}$ for unbinding as $T\rightarrow 0$ and no re-entrance at low activity. If $D_R^a$ is large enough then re-entrance is abolished altogether.
For $|v|>|v_c(T)|$, the effective pair potential $\mathcal{U}_{\mathrm{eff}}$ develops a maximum as in Fig.~\ref{fig:strawman}, thereby implying that incipient defect pairs unbind for arbitrarily small temperature. The physical picture is then quite clear. At low activity, rotational diffusion randomizes the orientation of the $+1/2$ disclination and makes its motion less persistent, allowing the defect pair to remain bound. It is in this way that noise counterintuitively stabilizes the ordered nematic phase. At higher activity, the active torques compete with rotational diffusion, but ultimately enhance the persistent nature of defect motion. In this case rotational noise becomes irrelevant and we recover the scenario sketched in Fig.~\ref{fig:strawman}. The simple one-dimensional model predicts defect unbinding self-consistently if the persistence length of the $+1/2$ disclination ($|v|/D_R$) is greater than the position of the barrier in the potential ($r_c=K/(|v|\gamma)$). Equating the two lengths, we obtain the same threshold scaling as in Eq.~\ref{eq:v0c} at low $T$. We have verified this scenario by numerically integrating Eqs.~\ref{eq:eom} and~\ref{eq:edot} for either sign of $v$,
as shown in Fig.~\ref{fig:dist}.

In summary, starting from the equations of motion of a two-dimensional active nematic, we have derived the statistical dynamics of its topological defects as a noisy  mixture of motile and non-motile particles.
Through a Fokker-Planck approach, we show perturbatively that the rotational diffusion of $+1/2$ defects allows the nematic phase to survive defect proliferation below an activity threshold. We identify, for small activity, the temperature-activity locus of a BKT-like active-nematic/isotropic transition, and provide arguments suggesting that defects are unbound at any nonzero temperature above a critical activity, and that a re-entrant disordered phase arises at low temperature.
Venturing beyond the present perturbative approach and taking many-defect features, such as screening, into account are clearly the immediate challenges.

We thank Ananyo Maitra and Mike Cates for insightful comments.
This work was supported by the National Science Foundation at Syracuse University through awards DMR-1609208 (MCM \& SS) and DGE-1068780 (MCM) and at KITP under Grant PHY-1748958. MCM, MJB \& SS thank the Syracuse Soft and Living Matter Program for support. All authors thank the Simons Foundation for support and the KITP for hospitality in the course of this work. SR acknowledges the support of the Tata Education and Development Trust, and a J C Bose Fellowship of the Science \& Engineering Research Board, India.




\pagebreak
\widetext
\begin{center}
\textbf{\large Defect Unbinding in Active Nematics\\~\\
SUPPLEMENTARY INFORMATION\\}
	\vspace{1.3em}
	Suraj Shankar$^{a,b}$, Sriram Ramaswamy$^{b,c}$, M. Cristina Marchetti$^{a,b}$ and Mark J. Bowick$^{a,b}$\\
	\textit{$^a$Physics Department and Syracuse Soft and Living Matter Program,\\ Syracuse University, Syracuse, NY 13244, USA.\\
$^b$Kavli Institute for Theoretical Physics, University of California, Santa Barbara, CA 93106, USA.\\
$^c$Centre for Condensed Matter Theory, Department of Physics,\\ Indian Institute of Science, Bangalore 560 012, India.}
\end{center}

\setcounter{equation}{0}
\setcounter{figure}{0}
\setcounter{table}{0}
\setcounter{page}{1}
\makeatletter
\renewcommand{\theequation}{S\arabic{equation}}
\renewcommand{\thefigure}{S\arabic{figure}}
\renewcommand{\bibnumfmt}[1]{[S#1]}
\renewcommand{\citenumfont}[1]{S#1}
\thispagestyle{empty}

\section{I.\hspace{1em} Derivation of equations of motion for $\pm1/2$ disclinations}
\label{sec:defecteom}
Consider the following general continuum equations for the hydrodynamics of a nematic liquid crystal written in terms of the tensor order parameter $\b{Q}=S(2\hat{\b{n}}\hat{\b{n}}-\b{1})$ and the flow velocity $\b{u}$.
\begin{gather}
	\partial_t\b{Q}+\b{u}\cdot{\bm\nabla}\b{Q}-[\b{Q},\Omega]=\b{L}(\b{Q},\b{u})+\dfrac{1}{\gamma}\left[a_2-a_4\ \tr(\b{Q}^2)\right]\b{Q}+\dfrac{K}{\gamma}\del^2\b{Q}\ ,\label{eq:Qeqn}\\
	\b{L}(\b{Q},\b{u})=\lambda_1\b{D}+\lambda_2\b{Q}{\bm\nabla}\cdot\b{u}-\lambda_3\b{Q}\mathrm{tr}(\b{Q}\cdot{\bm\nabla}\b{u})\ ,\nonumber
\end{gather}
where $[A,B]=AB-BA$ is the matrix commutator, $2\Omega_{\mu\nu}=\del_{\mu}u_{\nu}-\del_{\nu}u_{\mu}$ is the vorticity tensor and $2D_{\mu\nu}=\del_{\mu}u_{\nu}+\del_{\nu}u_{\mu}-\delta_{\mu\nu}{\bm\nabla}\cdot\b{u}$ is the symmetrized and traceless strain rate tensor. We work in the one Frank elastic constant ($K$) approximation, but have included various flow alignment terms ($\lambda_{1,2,3}$) in $\b{L}(\b{Q},\b{u})$. Since the system is in contact with a substrate, we neglect inertia and determine the flow velocity through the Stokes equation that balances friction and active stresses as
\begin{equation}
	\Gamma\b{u}=\del\cdot\vec{\sigma}^a\;,
	\label{eq:Stokes}
\end{equation}
where $\Gamma$ is the friction with the substrate and $\vec{\sigma}^a=\alpha\b{Q}$ is the active stress. We neglect the passive elastic and viscous stresses as they are all higher order in gradients. We assume the nematic density to be constant even though, in distinction from previous work \cite{SIpismen2013dynamics,SIgiomi2014defect}, we do not consider the flow to be incompressible ($\del\cdot\b{u}\neq 0$). We will explore elsewhere the effect of incompressibility or a density field with conserving dynamics, but do not expect these will introduce drastic differences with respect to the behavior described here.
We assume we are deep in the nematic state where $a_2>0$.

We rescale $\b{r}\to\b{r}/\xi$, $t\to t/\tau$ and $\b{Q}\to\b{Q}/S_{\mathrm{max}}$, where
\begin{equation}
	\xi=\sqrt{\dfrac{K}{a_2}}\ ,\quad\tau=\dfrac{\gamma}{a_2}\quad\mathrm{and}\quad S_{\mathrm{max}}=\sqrt{\dfrac{a_2}{2a_4}}\ .\label{eq:scales}
\end{equation}
	The active stress is also non-dimensionalized as $\bar{\alpha}=\alpha\tau S_{\mathrm{max}}/(\Gamma\xi^2)$. The order parameter $\b{Q}$ is a rank-$2$ symmetric and traceless tensor. Hence it can be rewritten in terms of a single complex variable $\chi=Q_{xx}+iQ_{xy}=Se^{2i\theta}$, where $\theta$ is the angle of the director $\hat{\b{n}}=\left(\cos\theta,\sin\theta\right)$. In terms of $\chi$, Eq.~\ref{eq:Qeqn} now becomes (in dimensionless form)
\begin{equation}
	\partial_t\chi+\b{u}\cdot{\bm\nabla}\chi-i({\bm\del}\times\b{u})\chi=(\bar{\lambda}_1-\bar{\lambda}_3|\chi|^2)D-\bar{\lambda}_3\chi^2 D^*+\lambda_2\chi({\bm\nabla}\cdot\b{u})+(1-|\chi|^2)\chi+\del^2\chi\ ,
\end{equation}
where $D=D_{xx}+iD_{xy}=[\partial_{x}u_x-\partial_yu_y+i(\partial_xu_y+\partial_yu_x)]/2$ is the strain rate tensor $\b{D}$ in complex notation. Additionally, we have rescaled the flow alignment parameters as $\bar{\lambda}_{1,3}=\lambda_{1,3}/2S_{\mathrm{max}}$.

	Finally, we boost to a frame which is comoving and corotating with an isolated slowly moving $\pm1/2$ disclination. To do this, we perform the following coordinate and field transformations,
\begin{equation}
	\b{r}=\b{R}(t)\cdot\b{r}'(t)+\b{b}(t)\ ,\quad\b{u}(\b{r},t)=\b{R}(t)\cdot\b{u}'(\b{r}'(t),t)\ ,\quad\b{Q}(\b{r},t)=\b{R}(t)\cdot\b{Q}'(\b{r}'(t),t)\cdot\b{R}(t)^T\ ,
\end{equation}
	where $\b{R}(t)=\b{1}\cos\omega t+\vec{\epsilon}\sin\omega t$ with $\vec{\epsilon}$ as the Levi-Civita tensor is the rotation matrix rotating the frame counter-clockwise at a constant rate of $\omega$ ($\dot{\b{R}}(t)=\omega\vec{\epsilon}\cdot\b{R}(t)$ and $\b{R}^T\cdot\b{R}=\b{1}$), which is taken to be the angular rotation rate of the defect. The translational shift is determined to be $\b{b}(t)=-\b{R}(t)\cdot\vec{\epsilon}\cdot\b{v}/\omega$ by requiring $\dot{\b{r}}'(t)=-\omega\vec{\epsilon}\cdot\b{r}'(t)-\b{v}$, where $\b{v}$ is the constant translational velocity of the defect. The same transformation in terms of the complex field $\chi$ is given as
\begin{equation}
	\chi(\b{r},t)=\chi'(\b{r}'(t),t)e^{-2i\omega t}\ .
\end{equation}
	Using this, we obtain $\partial_t\chi=e^{-2i\omega t}[\partial_t\chi'-\b{v}\cdot\bm\del'\chi'+\omega\b{r}'\times\bm\del'\chi'-2i\omega\chi']$ and $\bm\del\chi=e^{-2i\omega t}\b{R}\cdot\bm\del'\chi'$ and similar expressions involving $\b{u}$.
	Assuming the dynamics is quasistatic in this moving frame of reference ($\partial_t\chi'=0$), we neglect any explicit time dependence and drop the primes on all the variables to get
\begin{equation}
	(-\b{v}+\b{u})\cdot{\bm\nabla}\chi+\omega\b{r}\times{\bm\nabla}\chi-i(2\omega+\del\times\b{u})\chi=(\bar{\lambda}_1-\bar{\lambda}_3|\chi|^2)D-\bar{\lambda}_3\chi^2 D^*+\lambda_2\chi({\bm\nabla}\cdot\b{u})+(1-|\chi|^2)\chi+\del^2\chi\ ,\label{eq:fulleqn}
\end{equation}
where $\b{v}$ and $\omega$ are the velocity and angular rotation rate of the $\pm1/2$ disclination.
	Assuming the defect is moving and rotating slowly, and the activity is also small, we take $\b{v},\omega,\bar{\alpha}=\mathcal{O}(\eta)$, where $\eta\ll 1$ will be used as a book-keeping parameter to do perturbation theory in. This allows us to treat all the backflow terms perturbatively. Expanding $\chi=\chi_0+\eta\chi_1+\mathcal{O}(\eta^2)$ and $\b{u}=\b{u}_0+\eta\b{u}_1+\mathcal{O}(\eta^2)$, we have at $\mathcal{O}(\eta^0)$,
\begin{equation}
	\b{u}_0=\b{0}\ ,\quad\mathrm{and}\quad\del^2\chi_0+(1-|\chi_0|^2)\chi_0=0\ .
\end{equation}
	This is the static equilibrium equation which is easily solved to obtain the configuration of a single stationary disclination. For a $\pm1/2$ disclination, we have $\chi_0(\b{r})=S_0(r)e^{\pm i\phi+2i\theta_0}$, where $r,\phi$ are standard polar coordinates in the plane with the defect centered at the origin and $\theta_0$ is a constant related to the orientation of the $\pm1/2$ disclination. $S_0(r)$ satisfies the following differential equation
\begin{equation}
	S_0''+\dfrac{S_0'}{r}+\left(1-\dfrac{1}{r^2}-S_0^2\right)S_0=0\ .
\end{equation}
	whose solution can be written in terms of a  Pad{\'e} approximant (with upto 3\% error) with the correct asymptotic behavior ($S_0\simeq1-1/2r^2$ as $r\to\infty$ and $S_0\simeq0.5831 r$ as $r\to 0$), given by \cite{SIpismen1999vortices}
\begin{equation}
	S_0(r)=r\sqrt{\dfrac{0.34+0.07r^2}{1+0.41r^2+0.07r^4}}\ .
\end{equation}
	We shall use this approximate form of $S_0(r)$ later on to compute the defect mobility.
	
The equations for the first order corrections $\chi_1(\b{r})$ and $\b{u}_1(\b{r})$ are obtained by retaining terms of $\mathcal{O}(\eta)$ in Eqs.~(\ref{eq:fulleqn}) and (\ref{eq:Stokes}),
\begin{gather}
	\del^2\chi_1+(1-2|\chi_0|^2)\chi_1-\chi_0^2\chi^*_1=\mathcal{I}(\b{r})\label{eq:chi1}\\
	\mathcal{I}(\b{r})=(\b{u}_1-\b{v})\cdot{\bm\nabla}\chi_0-i(2\omega+{\bm\nabla}\times\b{u}_1)\chi_0+\omega\b{r}\times{\bm\nabla}\chi_0-(\bar{\lambda}_1-\bar{\lambda}_3|\chi_0|^2)D_1+\bar{\lambda}_3\chi_0^2D^*_1-\lambda_2\chi_0({\bm\nabla}\cdot\b{u}_1)\label{eq:I}\\
	\b{u}_1=\bar\alpha{\bm\nabla}\cdot\b{Q}_0\ ,
\end{gather}
where $\b{Q}_0$ is the tensor form of $\chi_0$ corresponding to the static defect solution and $D_1$ is the corresponding complex deviatoric strain rate computed using $\b{u}_1$. The first order velocity correction is easily determined to be
\begin{equation}
	\b{u}_1=\bar{\alpha}\left(S_0'(r)+\dfrac{S_0(r)}{r}\right)\left(\cos2\theta_0,\sin2\theta_0\right)\ .\label{eq:u1}
\end{equation}
The corresponding first order corrections to the vorticity ($\bm\del\times\b{u}$), divergence ($\bm\del\cdot\b{u}$) and deviatoric shear in complex form ($D$) are also easily obtained to be
\begin{gather}
	\bm\del\times\b{u}_1=\bar{\alpha}\sin(\phi-2\theta_0)S_0(r)\left[1-S_0(r)^2\right]\ ,\label{eq:du1}\\
	\bm\del\cdot\b{u}_1=-\bar{\alpha}\cos(\phi-2\theta_0)S_0(r)\left[1-S_0(r)^2\right]\ ,\label{eq:du2}\\
	D=-\bar{\alpha}S_0(r)\left[1-S_0(r)^2\right]\ e^{i(\phi+2\theta_0)}\ .\label{eq:du3}
\end{gather}
These terms contribute to the inhomogeneity $\mathcal{I}(\b{r})$ as shown in Eq.~\ref{eq:chi1},\ref{eq:I}.
The linear complex differential operator on the left-hand side of Eq.~\ref{eq:chi1} given by
\begin{equation}
	\mathcal{L}(f,f^*)=\del^2f+(1-2|\chi_0|^2)f-\chi_0^2f^*\ ,
\end{equation}
has three zero modes, denoted by $\Psi$: two corresponding to translations ($\Psi=\del\chi_0$) and one for rotation ($\Psi=i\chi_0$) of the defect. Multiplying Eq.~\ref{eq:chi1} by a complex conjugated zero mode $\Psi^*$ and integrating by parts, we get the following solvability condition for $\chi_1$ (Fredholm alternative)
\begin{equation}
	\mathrm{Re}\left(\int_{\Sigma}\dd^2r\ \mathcal{I}(\b{r})\Psi^*(\b{r})\right)=\mathrm{Re}\left(\oint_{\partial\Sigma}\left(\Psi^*\partial_{\perp}\chi_1-\chi_1\partial_{\perp}\Psi^*\right)\right)\ ,\label{eq:fredholm}
\end{equation}
where $\Sigma$ is any region of space and $\partial_{\perp}$ is the normal derivative along the boundary $\partial\Sigma$. We shall apply this solvability condition on a circular region of radius $r_0=\mathcal{O}(\eta^{-1/2})$. This choice of $r_0$ is motivated by the fact that the perturbative expansion of $\chi$ breaks down in the very far-field ($r\sim \eta^{-1}$), while a pure phase analysis performed later is valid on scales all the way from infinity down to $r\sim\eta^{-1/2}$. So for $r_0=\mathcal{O}(\eta^{-1/2})$, both inner and outer expansion can be asymptotically matched.

Projecting onto the rotational zero mode ($\Psi=i\chi_0$), we find that all the terms involving the flow (independent of $\omega$) in the $\mathcal{I}$ integral have a vanishing real part as required in the left hand side of Eq.~\ref{eq:fredholm} and so $\omega=0$ at this order. Using the translational eigenmodes ($\Psi=\del\chi_0$), as $\b{u}\sim 1/r$ for $r\to\infty$, all the integrals involving the flow velocity are convergent for large $r$ (so we extend the integral to infinity), while only the term involving the defect velocity $\b{v}$ is log-divergent. So we have for the $+1/2$ disclination
\begin{equation}
	\mathrm{Re}\int^{r_0}_0\dd^2r\mathcal{I}(\b{r}){\bm\nabla}\chi_0^*=\bar{\alpha}\ \hat{\b{e}}_0(5.735+0.6284\lambda_2)-\pi\b{v}\ln\left(\dfrac{r_0}{1.126}\right)\ ,
\end{equation}
where $\hat{\b{e}}_0=(\cos2\theta_0,\sin2\theta_0)$ is the unit polarization of the $+1/2$ disclination. Note that, as expected by symmetry, the active backflow terms don't contribute for the $-1/2$ disclination (the $\bar{\alpha}$ term is absent). Interestingly, we find that both $\bar{\lambda}_1$ and $\bar{\lambda}_3$ don't contribute to the mobility calculation at this order, though they do enter the linear stability analysis of fluctuations about the homogeneous ordered state \cite{SIsrivastava2016negative}.

For the line integral on the RHS of Eq.~\ref{eq:fredholm}, we have to evaluate $\chi_1$ at $r_0=\mathcal{O}(\eta^{-1/2})$. For arbitrary $\b{v}$ and $\bar\alpha$ ($\mathcal{O}(\eta^0)$), we now have for large $r_0$, $S=1-\mathcal{O}(\eta)$ and ${\bm\del}\b{u}_1=\mathcal{O}(\eta)$ on the scale of $r_0$. So in the very far-field, we set $S=1$ and have only the phase $\theta$ at leading order,
\begin{equation}
	\b{v}\cdot{\bm\nabla}\theta+\del^2\theta=0\ .
\end{equation}
This equation, when solved with the correct winding condition on $\theta$ ($\oint\dd\theta=\pm\pi$ around the origin for the $\pm1/2$ defect), gives rise to a highly anisotropic phase profile given by \cite{SIpismen1999vortices}
\begin{gather}
	{\bm\nabla}\theta=\pm\dfrac{1}{2}e^{-\b{v}\cdot\b{r}/2}\vec{\epsilon}\cdot\left[\b{v}K_0\left(\dfrac{|\b{v}|r}{2}\right)-|\b{v}|\hat{\b{r}}K_1\left(\dfrac{|\b{v}|r}{2}\right)\right]+\vec{\epsilon}\cdot\b{v}A\ ,
\end{gather}
where $\vec{\epsilon}$ is the antisymmetric Levi-Civita tensor; $K_{0}(x)$, $K_{1}(x)$ are modified Bessel functions of the first kind and $A$ is a constant that we will later relate to a weak external phase gradient. This gives the far field phase solution for both $+1/2$ and $-1/2$ defects when moving. Expanding this outer solution for small $\b{r}$ to match to the inner solution at $r\sim r_0$, we use it in the line integral in Eq.~\ref{eq:fredholm} to get
\begin{equation}
	\mathrm{Re}\int_0^{2\pi}\dd\phi\ r_0\left({\bm\nabla}\chi_0^*\partial_r\chi_1-\chi_1\partial_r{\bm\nabla}\chi_0^*\right)|_{r=r_0}=-\pi\b{v}\ln\left(\dfrac{|\b{v}|r_0}{4}e^{\gamma_E-1/2\mp2A}\right)\ ,
\end{equation}
where $\gamma_E$ is the Euler-Mascheroni number. Putting all this together into Eq.~\ref{eq:fredholm} and writing $\b{v}A=-\vec{\epsilon}\cdot{\bm\nabla}\theta_{\mathrm{ext}}$ (a weak external phase gradient orthogonal to the defect motion), the $r_0$ dependence cancels out leaving us with the following defect mobility relation for both $+1/2$ and $-1/2$ disclinations (denoted using $\pm$ correspondingly),
\begin{equation}
	\b{v}_{\pm}\ln\left(\dfrac{3.29}{|\b{v}_{\pm}|}\right)=\pm 2\vec{\epsilon}\cdot{\bm\nabla}\theta_{\mathrm{ext}}+\delta_{q_i,1/2}\ \bar{\alpha}\ \hat{\b{e}}_0(5.735+0.6284\lambda_2)\ .\label{eq:mobility}
\end{equation}
The $-1/2$ disclination unlike the $+1/2$ disclination, moves only under the influence of passive phase gradients through a Magnus like force and the active self-induced flow vanishes. Also note that, the numerical prefactors present in Eq.~\ref{eq:mobility} are not universal (though the structure of the equation is) and they can change by using a different model for the profile of the defect core. One can restore dimensions appropriately using the length and times scales used previously (see Eq.~\ref{eq:scales}). In the presence of other distant defects, ${\bm\nabla}\theta_{\mathrm{ext}}$ is given by the cumulative phase gradient at the core of the considered defect due to the presence of other defects, which in the pure phase approximation can be obtained by linear superposition as
\begin{equation}
	{\bm\nabla}\theta_{\mathrm{ext}}(\b{r}_i)=-\vec{\epsilon}\cdot{\bm\nabla}_{\b{r}_i}\sum_{j\neq i}q_j\ln\left|\dfrac{\b{r}_i-\b{r}_j}{a}\right|\ ,\label{eq:gradtheta}
\end{equation}
where $q_j=\pm1/2$ is the charge of the $j^{\mathrm{th}}$ defect and $a\sim\xi$ is the defect core size that functions as a microscopic cutoff. The motion of the defect transverse to the phase gradient at its core is a manifestation of the equilibrium Magnus like force, which when eventually written in terms of the defect positions using Eq.~\ref{eq:gradtheta}, exactly corresponds to a force arising from the equilibrium Coulombic interaction between defects. In general, we find that the mobility relation is nonlinear with a logarithmic correction depending on the speed of the defect. As this correction is rather weak, to simplify matters, we use a constant mobility in the main text which should be valid for slowly moving defects or when the defect density is large enough that the interactions are screened.

Note that the above calculation still misses some features of active defect motion. As can be seen from Eq.~\ref{eq:u1}, though the velocity field is frictionally screened, it decays slowly as $\b{u}_1\sim 1/r$. Hence, along with the self-induced active backflow of the $+1/2$ defect that leads to its spontaneous motility, one must also include in $\b{u}_1$ the instantaneous active flow generated by all other distant $\pm1/2$ defects, as this flow is of the same order as the phase gradient that generates the equilibrium passive defect interaction (Eq.~\ref{eq:gradtheta}). One can write $\b{u}_1=\bar{\alpha}\bm\del\b{Q}_0+\tilde{\b{u}}_1$, where $\b{Q}_0$ is the defect configuration for the $i^{\mathrm{th}}$ defect and $\tilde{\b{u}}_1$ includes the entrainment of the $i^{\mathrm{th}}$ defect by the flow of other defects. A full calculation of $\tilde{\b{u}}_1$ including many defects is not analytically tractable as of now, but for weak phase gradients if we approximately linearize $\b{Q}$ assuming $S\simeq 1$ to be constant in the far-field, then the phase due to various defects just adds up linearly. In this case, $\tilde{\b{u}}_1\simeq2\bar{\alpha}(\partial_y\theta,\partial_x\theta)$ and using Eq.~\ref{eq:gradtheta} for the phase gradient gives
\begin{equation}
	\tilde{\b{u}}_1(\b{r})=2\bar{\alpha}\vec{\sigma}_3\cdot\bm\del\sum_{j\neq i}q_j\ln\left|\dfrac{\b{r}-\b{r}_j}{a}\right|\ ,
	\label{eq:u1tilde}
\end{equation}
where $\vec{\sigma}_3$ is the third Pauli matrix (so $\vec{\sigma}_3\cdot\bm\del=(\partial_x,-\partial_y)$). Including a ``second'' active coefficient \cite{SImaitra2018nonequilibrium}, we have an additional force contribution to the right hand side of the flow equation (Eq.~\ref{eq:Stokes}) $\sim \alpha'\b{Q}\cdot(\bm\del\cdot\b{Q})$. This term does not affect the self-propulsion of a $+1/2$ defect, but it does contribute a term $\sim\alpha'\bm\del\theta$ to the flow entrainment term $\tilde{\b{u}}_1$. As is evident, both active terms are comparable to the passive elastic one and in general generate an anisotropic, linear in activity, correction to the Frank elastic constant. This is akin to effectively having a different bend ($K_3$) and splay ($K_1$) elastic constant. Depending on where one is in parameter space, the linear active corrections to nematic elasticity may be stabilizing or destabilizing in nature, recovering the same effect obtained in a linear stability calculation about the homogeneous ordered state \cite{SIsrivastava2016negative,SImaitra2018nonequilibrium}. In all of our work, we shall remian in the regime where both $K_1$ and $K_3$ are always positive even after including the linear activity corrections, allowing for a stable base nematic. Additionally, upon including fluctuations in a homogeneous active nematic, following the arguments in Ref.~\cite{SIshankar2018low}, the $\bar{\alpha}$ contribution to $\tilde{\b{u}}_1$ in Eq.~\ref{eq:u1tilde} being anisotropic, is irrelevant on large scales, hence we neglect it in our calculations and discussion in the main text.

\section{II.\hspace{1em} Derivation of polarization dynamics of the $+1/2$ disclination}
Here we obtain the dynamics of the polarization vector $\b{e}_i(t)=a{\bm\del}\cdot\b{Q}(\b{r}^+_i(t))$ for the $i^{\mathrm{th}}$ $+1/2$ defect. Let $\b{E}(\b{r},t)=a{\bm\del}\cdot\b{Q}$. Then $\b{e}_i(t)=\b{E}(\b{r}^+_i(t),t)$. Taking a time derivative we get
\begin{equation}
	\dfrac{\dd\b{e}_i(t)}{\dd t}=\partial_t\b{E}(\b{r}^+_i,t)+\dfrac{\dd\b{r}^+_i}{\dd t}\cdot{\bm\del}_{\b{r}^+_i}\b{E}(\b{r}^+_i,t)\ .\label{eq:edotdef}
\end{equation}
From the previous section (Sec.~I), we found the self-induced rotation of the $+1/2$ defect to be zero ($\omega=0$) at the order we are working, so we do not include it here and don't consider it further. This simplifies things greatly as it allows us to set the time-dependent rotation matrix to the constant unit tensor ($\b{R}=\b{1}$). As $\partial_t\b{E}=\b{0}$ in the quasistatic approximation, we only have the second term. We shall address the implications of relaxing this upon including smooth director distortions in the next section (Sec.~III). Writing $\b{v}_i(t)=\dot{\b{r}}^+_i(t)$ (the $+1/2$ defect velocity) and using the definition of $\b{E}$, we have
\begin{equation}
	\dfrac{\dd\b{e}_i(t)}{\dd t}=a\b{v}_i(t)\cdot\left(\lim_{\b{r}\to\b{r}_i(t)}{\bm\del}{\bm\del}\cdot\b{Q}(\b{r},t)\right)\ .\label{eq:dedt}
\end{equation}
Computing ${\bm\del\bm\del}\cdot\b{Q}$ perturbatively about the stationary equilibrium defect, we expand Eq.~\ref{eq:fulleqn} as before to $\mathcal{O}(\eta)$ (where $\b{v},\bar{\alpha}\sim\eta$). As we eventually have to take the $\b{r}\to\b{r}_i(t)$ limit (at the center of the defect core), we find it easier to expand in $S$ and $\theta$ instead of the complex variable $\chi$. Using the same notation as before and setting $\omega=0$, Eq.~\ref{eq:fulleqn} in terms of $S$ and $\theta$ is given as
\begin{gather}
	(\b{u}-\b{v})\cdot\bm\del S=(\bar{\lambda}_1-\bar{\lambda}_3S^2)\mathrm{Re}\left\{e^{-2i\theta}D\right\}-\bar{\lambda}_3S^2\mathrm{Re}\left\{e^{2i\theta}D^*\right\}+\lambda_2S\bm\del\cdot\b{u}+(1-S^2)S+\del^2S-4S|\bm\del\theta|^2\ ,\\
	2(\b{u}-\b{v})\cdot\bm\del\theta-\bm\del\times\b{u}=\left(\dfrac{\bar{\lambda}_1}{S}-\bar{\lambda}_3S\right)\mathrm{Im}\left\{e^{-2i\theta}D\right\}-\bar{\lambda}_3S\ \mathrm{Im}\left\{e^{2i\theta}D^*\right\}+2\del^2\theta+\dfrac{2}{S}\bm\del S\cdot\bm\del\theta\ .
\end{gather}
We have dropped the subscript $i$ to keep the notation uncluttered. Writing $S=S_0(r)+\eta S_1+\mathcal{O}(\eta^2)$ and $2\theta=\phi+2\theta_0+\eta2\theta_1+\mathcal{O}(\eta^2)$ ($\theta_0$ is a constant setting the orientation of the $+1/2$ defect polarization), we use Eqs.~\ref{eq:du1},~\ref{eq:du2} and~\ref{eq:du3} to get the following linearized equations
\begin{gather}
	-\bar{\alpha}\left(S_0'+\dfrac{S_0}{r}\right)\dfrac{\sin(\phi-2\theta_0)}{r}+\dfrac{|\b{v}|\sin\phi}{r}+\bar{\alpha}\sin(\phi-2\theta_0)\left(S_0''+\dfrac{S_0'}{r}-\dfrac{S_0}{r^2}\right)=2\del^2\theta_1+\dfrac{2}{S_0}S_0'\partial_r\theta_1+\dfrac{1}{S_0}\dfrac{\partial_{\phi}S_1}{r^2}\ ,\\
	\left(1-3S_0^2-\dfrac{1}{r^2}\right)S_1+\del^2S_1-\dfrac{2S_0}{r^2}\partial_{\phi}\theta_1=\bar{\alpha}\left(S_0'+\dfrac{S_0}{r}\right)S_0'\cos(\phi-2\theta_0)-|\b{v}|S_0'\cos\phi\nonumber\\
		\qquad\qquad\qquad\qquad\qquad\qquad\qquad\qquad\qquad\qquad-\bar{\alpha}\left[\bar{\lambda}_1-2\bar{\lambda}_3S_0^2+\lambda_2S_0\cos(\phi-2\theta_0)\right]\left(S_0''+\dfrac{S_0'}{r}-\dfrac{S_0}{r^2}\right)\ .
\end{gather}
Here the defect is centered at the origin and $\b{v}$ has been taken along the $x$-axis without loss of generality. Note that for $\theta_0\neq0,\pi$, the defect polarization $\b{e}$ is not aligned with its velocity $\b{v}$. We note, by inspection that the following ansatz solves the above equations,
\begin{gather}
	S_1=\psi_0(r)+\psi_1(r)\cos\phi+\psi_2(r)\cos(\phi-2\theta_0)\ ,\\
	\theta_1=\Theta_1(r)\sin\phi+\Theta_2(r)\sin(\phi-2\theta_0)\ .\label{eq:thetaexp}
\end{gather}
 The first order corrections ($\psi_{0,1,2}(r)$ and $\Theta_{1,2}(r)$) satisfy the following equations.
\begin{gather}
	2\left(\Theta_1''+\dfrac{\Theta_1'}{r}-\dfrac{\Theta_1}{r^2}\right)+\dfrac{2S_0'}{S_0}\Theta_1'-\dfrac{\psi_1}{S_0r^2}-\dfrac{|\b{v}|}{r}=0\ ,\label{eq:eq0}\\
	\bar{\alpha}\left[S_0(1-S_0^2)+\dfrac{1}{r}\left(S_0'+\dfrac{S_0}{r}\right)\right]+2\left(\Theta_2''+\dfrac{\Theta_2'}{r}-\dfrac{\Theta_2}{r^2}\right)+\dfrac{2S_0'}{S_0}\Theta_2'-\dfrac{\psi_2}{S_0r^2}=0\ ,\\
	\left(1-3S_0^2-\dfrac{1}{r^2}\right)\psi_0+\psi_0''+\dfrac{\psi_0'}{r}=\bar{\alpha}S_0(\bar{\lambda}_1-2\bar{\lambda}_3S_0^2)(1-S_0^2)\ ,\\
	\left(1-3S_0^2-\dfrac{2}{r^2}\right)\psi_1+\psi_1''+\dfrac{\psi_1'}{r}-\dfrac{2S_0}{r^2}\Theta_1+|\b{v}|S_0'=0\ ,\\
	\left(1-3S_0^2-\dfrac{2}{r^2}\right)\psi_2+\psi_2''+\dfrac{\psi_2'}{r}-\dfrac{2S_0}{r^2}\Theta_2=\bar{\alpha}\left[S_0'\left(S_0'+\dfrac{S_0}{r}\right)+\lambda_2S_0^2(1-S_0^2)\right]\ .\label{eq:eqn}
\end{gather}
Using the first order solution, we expand $\dd\b{e}/\dd t$ to $\mathcal{O}(\eta)$ (not counting the explicit factor of $\b{v}$ multiplying $\del\del\cdot\b{Q}$ in Eq.~\ref{eq:dedt}). Projecting $\dd\b{e}/\dd t$ along and transverse to $\b{e}$, we have
\begin{align}
	\dfrac{1}{|\b{e}|^2}\b{e}\cdot\dfrac{\dd\b{e}}{\dd t}&=\eta|\b{v}|\left[\dfrac{2S_0'(0)\Theta_1'(0)+3\psi_1''(0)}{4S_0'(0)}+\cos2\theta_0\left(\dfrac{2S_0'(0)\Theta_2'(0)+3\psi_2''(0)}{4S_0'(0)}\right)+\mathcal{O}(\eta)\right]\ ,\\
	\dfrac{1}{|\b{e}|^2}\b{e}\times\dfrac{\dd\b{e}}{\dd t}&=-\eta|\b{v}|\left(\sin2\theta_0\dfrac{6S_0'(0)\Theta_2'(0)+\psi_2''(0)}{4S_0'(0)}+\mathcal{O}(\eta)\right)\ .
\end{align}
As $r\to 0$ (at the core of the defect), the scalar order parameter must vanish ($S\to 0$). Taking the $r\to 0$ limit of the first order equations (Eqs.~\ref{eq:eq0}-\ref{eq:eqn}), we can deduce the leading behaviour of all the first order corrections for small $r$. Requiring that the physical solution not blow up at the core and $\chi$ be single valued, we obtain
\begin{gather}
	\psi_0(0)=\psi_1(0)=\psi_2(0)=0\ ,\quad\Theta_1(0)=\Theta_2(0)=0\ ,\\
	\psi_0'(0)=\bar{\lambda}_1\bar{\alpha}S'_0(0)\ ,\quad\psi_1'(0)=\psi_2'(0)=0\ ,\\
	\psi_1''(0)=-|\b{v}|S_0'(0)\ ,\quad\psi_2''(0)=2\bar{\alpha}S_0'(0)^2\ ,\quad\psi_0''(0)=0\ ,\\
	\Theta_1'(0)=\dfrac{|\b{v}|}{4}\ ,\quad\Theta_2'(0)=-\dfrac{\bar{\alpha}}{2}S_0'(0)\ .
\end{gather}
Using this, along with the fact that $|\b{e}|=2S_0'(0)+\mathcal{O}(\eta)$ ($a\sim\xi$) and writing $|\b{v}|\cos2\theta_0$, $|\b{v}|\sin2\theta_0$ in terms of scalar and cross products of $\b{v}$ and $\b{e}$, we obtain for the $i^{\mathrm{th}}$ $+1/2$ disclination (now setting the book-keeping parameter $\eta=1$)
\begin{equation}
	\dfrac{\dd\b{e}_i}{\dd t}=-\dfrac{5}{8}\b{v}_i\cdot\left(\b{v}_i-\bar{\alpha}\b{e}_i\right)\b{e}_i-\dfrac{\bar{\alpha}}{8}(\b{v}_i\times\b{e}_i)\vec{\epsilon}\cdot\b{e}_i\ ,\label{eq:edotdet}
\end{equation}
with $\b{v}_i=\dot{\b{r}}^+_i(t)$ as the $+1/2$ defect velocity. Restoring appropriate dimensional units, we obtain the equation quoted in the main text.

\section{III.\hspace{1em} Non-quasistatic solution: Rotational diffusion}
\label{sec:rotdiff}
In the previous section, the systematic part of the polarization dynamics was derived from the deterministic equations for the evolution of $\b{Q}$. In particular the equation was derived under the assumption of having a single isolate $+1/2$ defect in whose comoving and corotating frame, the dynamics is stationary. There are two effects that invalidate this assumption - the presence of noise which induces random fluctuations in the local frame and the contribution of the motion of other distant defects (over and above their effect included in $\b{v}_i$ through the passive Coulomb interaction). We shall include both here. As we interested in the term $\partial_t\b{E}$ (Eq.~\ref{eq:edotdef}) in the local defect frame, the main contribution comes from the slow relaxation of director distortions and not from the fast relaxation of the scalar order parameter. Working then in the pure-phase approximation, we have
\begin{equation}
	\theta(\b{r},t)=\sum_{i}q_i\phi(\b{r}-\b{r}_i(t))+\delta\theta(\b{r},t)+\theta_0\ ,\label{eq:thetanew}
\end{equation}
where $\theta_0$ is a constant, $\phi(\b{r})=\tan^{-1}(y/x)$ is the angle function and $q_i=\pm1/2$ are the defect strengths. The decomposition of $\theta(\b{r})$ is unambiguous if we demand $\delta\theta$ to be smooth and univalued everywhere so that $\bm\del\times\bm\del\delta\theta=0$. This captures all the smooth distortions of the director field present along with the topological defects. Deep in the ordered phase of a fluctuating $2d$ nematic liquid crystal (even when active), the smooth and single-valued director distortions analogous to ``spin-waves'' in magnets are a broken symmetry variable and so they are long-lived and easy to excite at low noise. These fluctuations cause random rotations of the $+1/2$ defect polarization. The smooth phase fluctuations in a $2d$ dry active nematic (number conserving \cite{SIshankar2018low} or not \cite{SImishra2010dynamic}) satisfy
\begin{equation}
	\partial_t\delta\theta=\dfrac{K}{\gamma}\del^2\delta\theta+f(\b{r},t)\ ,\label{eq:dthetasw}
\end{equation}
at long wavelengths, where $f(\b{r},t)$ is non-conserving white noise with zero mean and correlation $\langle f(\b{r},t)f(\b{r}',t')\rangle=2\Delta\delta(\b{r}-\b{r}')\delta(t-t')$. Using Eq.~\ref{eq:thetanew} we obtain for $\partial_t\b{E}(\b{r},t)=\partial_t\bm\del\cdot\b{Q}(\b{r},t)$ evaluated at the $i^{\mathrm{th}}$ defect core,
\begin{equation}
	\lim_{\b{r}\to\b{r}_i^+(t)}\partial_t\b{E}(\b{r},t)=-\vec{\epsilon}\cdot\b{e}_i\left(\omega_i+2\sum_{j\neq i}q_j\omega_j+2\partial_t\delta\theta(\b{r}_i^+)\right)\ ,
\end{equation}
where we have $\omega_j=\hat{\b{r}}_j\times\partial_t{\hat{\b{r}}}_j$ as the rotation speed of the $j^{\mathrm{th}}$ defect (both $\pm1/2$). Note that all the terms above being purely transverse, correspond to a torque on the polarization $\b{e}_i$. At leading order in perturbation theory, as shown in Sec.~I, we obtain the self-induced rotation $\omega_i=0$ for all the defects. That leaves us with the torque due to the elastic director distortions encoded in $\partial_t\delta\theta$. Note that specifying the positions and orientations of defects \emph{does not} uniquely define the smooth director distortion to interpolate between the defects. For a given configuration of defects and an \emph{independently} specified smooth director phase $\delta\theta$, one can solve Eq.~\ref{eq:dthetasw} setting the noise $f=0$, to obtain the torque experienced by the defects due to the particular imposed director configuration. This is precisely the computation performed by \citet{SIvromans2016orientational}, albeit numerically. \citet{SItang2017orientation} recognize the fact that for a given orientation of defects, there exist multiple interpolations of the smooth director field, all with differing energies. In the absence of noise ($f=0$), one can recover what both Refs.~\cite{SIvromans2016orientational,SItang2017orientation} call ``elastic torques'' by rephrasing Eq.~\ref{eq:dthetasw} as an initial value problem to compute the torque $2\partial_t\delta\theta$ for the given configuration.

In the presence of noise, all the smooth director distortions must be accounted for (even those with a higher energy), just with the appropriate statistical weight. Hence at steady state the relevant ``elastic'' contribution to the torque in a fluctuating description is actually stochastic and \emph{not} deterministic.
Using Eq.~\ref{eq:dthetasw}, we have
\begin{equation}
	\partial_t\delta\theta(\b{r}_i(t),t)=\int\dfrac{\dd^2q}{(2\pi)^2}\int\dfrac{\dd\omega}{2\pi}(-i\omega)\dfrac{f_{\b{q},\omega}\ e^{-i\omega t+i\b{q}\cdot\b{r}_i(t)}}{-i\omega+(K/\gamma)q^2}\ .
\end{equation}
One trivially has $\langle\partial_t\delta\theta(\b{r}_i)\rangle=0$ and evaluating its two-point correlation, we obtain
\begin{align}
	\langle\partial_t\delta\theta(\b{r}_i(t),t)\partial_{t'}\delta\theta(\b{r}_j(t'),t')\rangle&=2\Delta\int\dfrac{\dd^2q}{(2\pi)^2}\int\dfrac{\dd\omega}{2\pi}\dfrac{\omega^2}{\left[\omega^2+(K/\gamma)^2q^4\right]}e^{-i\omega(t-t')+i\b{q}\cdot[\b{r}_i(t)-\b{r}_j(t')]}\ ,\\
	&=2\Delta\left\{\delta(t-t')\dfrac{\delta_{ij}}{4\pi a^2}-\dfrac{K}{2\gamma}\int\dfrac{\dd^2q}{(2\pi)^2}\ q^2\ e^{-(K/\gamma)q^2|t-t'|+i\b{q}\cdot[\b{r}_i(t)-\b{r}_j(t')]}\right\}\\
	&=\dfrac{\Delta}{2\pi a^2}\left\{\delta(t-t')\delta_{ij}+a^2\dfrac{\gamma^2\left[r_{ij}^2-4(K/\gamma)|t-t'|\right]}{8K^2|t-t'|^3}e^{-\gamma r_{ij}^2/(4K|t-t'|)}\right\}\ ,\label{eq:swcorr}
\end{align}
where $r_{ij}=|\b{r}_i(t)-\b{r}_j(t')|$ ($r_{ij}\geq2a$ for $t=t'$, $i\neq j$) with $a$ as the defect core size. The random torque due to smooth director fluctuations has two parts, one that is delta-correlated and the other which retains memory of the past trajectory of defects and is exponentially small for well separated defects. Deep in the ordered phase, as $a$ is a microscopic length scale, the second term in Eq.~\ref{eq:swcorr} is negligible in comparison to the white noise part. Neglecting memory effects, we write $2\partial_t\delta\theta(\b{r}_i)=\sqrt{2D_R}\eta_i(t)$ and obtain the required rotational diffusion constant to be $D_R=\Delta/(\pi a^2)$. Neglecting active corrections to the noise, the translational noise in the defect dynamics is then $\mu T\sim\Delta$ and hence $D_R\sim\mu T/a^2$ which gives $\ell_R\sim a$ as stated in the main text.

\section{IV.\hspace{1em} Longitudinal noise in polarization dynamics}
Unlike the rotational noise that is more systematically derivable, as shown above, the longitudinal component of the noise $\vec{\nu}_i(t)$ is much harder to derive explicitly from noise in the $\b{Q}$ equation of motion. This noise generates fluctuations in the magnitude of the polarization $|\b{e}_i|$ and derives predominantly from noise in the scalar order parameter $S$. Within the low activity expansion, we assume the noise to the same as at equilibrium, in which case one can infer its properties by enforcing the fluctuation-dissipation theorem. In the absence of activity ($v=0$), our deterministic dynamics for $\b{e}$ (Eq.~\ref{eq:edotdet} in dimensionful form) reduces to
\begin{equation}
	\dot{\b{e}}_i=-\dfrac{5\mu^2\gamma}{8K}|\bm\del_i\mathcal{U}|^2\ \b{e}_i\ ,\label{eq:edoteq}
\end{equation}
where we have also used the fact that, in equilibrium $\b{v}_i=-\mu\bm\del\mathcal{U}$ is the deterministic part of either defect velocity. As the dynamics is overdamped and relaxational, both positional and orientational (polarization) dynamics must derive from a \emph{single} energy functional.

In the one-Frank constant approximation at equilibrium, spatial rotations and order parameter rotations get decoupled and become independent symmetries of the system. By virtue of this enhanced symmetry, the joint probability distribution of defect positions and polarizations must then decouple, with the positions governed by a Boltzmann weight with respect to the pair potential $\mathcal{U}$. This is also consistent with the derived deterministic dynamics, where the positional dynamics is independent of $\b{e}_i$ when $v=0$. Hence the $|\bm\del_i\mathcal{U}|^2$ term above in Eq.~\ref{eq:edoteq} is naturally viewed as part of a kinetic coefficient governing the relaxation of the polarization.

In order to really determine what constitutes the thermodynamic force in Eq.~\ref{eq:edoteq}, we need to write down the energy involving $\b{e}_i$. As $\b{e}_i=a\bm\del\cdot\b{Q}(\b{r}_i^+)$, to lowest order, the energetic contribution involving the polarization is $\sim |\b{e}_i|^2$. This term penalizes gradients of the alignment tensor as is expected of the elasticity of a nematic liquid crystal. Note that, a more complicated symmetry breaking mexican hat like potential for $\b{e}_i$ is unnatural as it favours the spontaneous creation of gradients and modulation of nematic order. Including higher order even polynomials is possible but it complicates the noise statistics by making it multiplicative. The simplest choice that retains just additive noise is the quadratic energy, which is what we choose. The polarization $\b{e}_i$ itself being dimensionless, we use $K$ for units of energy (the only thermodynamic parameter available). With this in place, we write the energy as $K|\b{e}_i|^2/2$, where the factor of two we have chosen arbitrarily (fixing an overall scale for $\b{e}_i$) for convenience. In this form, $\b{e}_i$ being an internal degree of freedom that depends on the defect core structure, the energy also represents the simplest contribution to the core energy.

Putting it all together, we obtain the joint distribution of defect positions and polarizations at equilibrium to be
\begin{equation}
	P_{\mathrm{eq}}^{2N}=\dfrac{1}{Z_{2N}}e^{-\mathcal{U}/T}\prod_{i=1}^N\left(\dfrac{ K}{2\pi T}e^{-K|\b{e}_i|^2/2T}\right)\ , \label{eq:defectBoltz}
\end{equation}
as quoted in the main text. This fixes the kinetic coefficient in Eq.~\ref{eq:edoteq} to be $(5\mu^2\gamma/8K^2)|\bm\del_i\mathcal{U}|^2$ which upon multiplying with $2T$ gives the two point correlation of the longitudinal noise $\vec{\nu}_i(t)$.

\section{V.\hspace{1em} Neutral $\pm1/2$ defect pair : Fokker-Planck Solution}
Consider an isolated $\pm1/2$ neutral defect pair in an active nematic. The dynamics of their pair separation $\b{r}=\b{r}^+-\b{r}^-$ and the $+1/2$ polarization $\b{e}$ is given by
\begin{gather}
	\dot{\b{r}}=v\b{e}-2\mu\bm\del\mathcal{U}+\sqrt{4\mu T}\vec{\xi}(t)\ ,\\
	\dot{\b{e}}=-\dfrac{\mu\gamma}{K}\vec{\zeta}\cdot\left[\mu\b{e}|\bm\del\mathcal{U}|^2-v|\b{e}|^2\bm\del\mathcal{U}\right]-\sqrt{2D_R}\ \vec{\epsilon}\cdot\b{e}\eta(t)+\vec{\nu}(t)\ ,
\end{gather}
where $\vec{\zeta}=(\b{1}+4\hat{\b{e}}\hat{\b{e}})/8$ and $\mathcal{U}(\b{r})=(\pi K/2)\ln(r/a)$. The corresponding Fokker-Planck equation is given by
\begin{align}
	\partial_tP+v\b{e}\cdot\bm\del P+&\dfrac{v\mu\gamma}{2K}\bm\del_{\b{e}}\cdot\left[(\b{e}\cdot\bm\del\mathcal{U})\b{e}P+\dfrac{1}{4}|\b{e}|^2\bm\del\mathcal{U}P\right]\nonumber\\
	&\quad=D_R(\b{e}\times\bm\del_{\b{e}})^2P+2\mu\bm\del\cdot\left[P\bm\del\mathcal{U}+T\bm\del P\right]+\dfrac{5\mu^2\gamma|\bm\del\mathcal{U}|^2}{8K^2}\bm\del_{\b{e}}\cdot\left[K\b{e}P+T\bm\del_{\b{e}}P\right]\ .
\end{align}
Taking moments of the polarization, $\rho(\b{r})=\int\dd^2e\ P(\b{r},\b{e})$, $\b{w}(\b{r})=\int\dd^2e\ \b{e}P(\b{r},\b{e})$ and $\b{M}(\b{r})=\int\dd^2e\ \b{e}\b{e}P(\b{r},\b{e})$, we have
\begin{gather}
	\partial_t\rho+v\bm\del\cdot\b{w}=2\mu\bm\del\cdot\left[\rho\bm\del\mathcal{U}+T\bm\del\rho\right]\ ,\label{eq:rhoeq}\\
	\partial_t\b{w}+v\bm\del\cdot\b{M}-\dfrac{v\mu\gamma}{2K}\left[\b{M}\cdot\bm\del\mathcal{U}+\dfrac{\tr(\b{M})}{4}\bm\del\mathcal{U}\right]=-\left(D_R+\dfrac{5\mu^2\gamma|\bm\del\mathcal{U}|^2}{8K}\right)\b{w}+2\mu\bm\del\cdot\left[\bm\del\mathcal{U}\b{w}+T\bm\del\b{w}\right]\ ,\label{eq:weq}\\
	\partial_t\b{M}=-4D_R\left(\b{M}+\tr(\b{M})\dfrac{\b{1}}{2}\right)+2\mu\bm\del\cdot(\bm\del\mathcal{U}\b{M}+T\bm\del\b{M})-\dfrac{5\mu^2\gamma|\bm\del\mathcal{U}|^2}{4K^2}(K\b{M}-T\rho\b{1})\ . \label{eq:Meq}
\end{gather}
where we have neglected all three and higher point correlations of the polarization and set $\langle\b{e}\b{e}\b{e}\rangle\simeq \b{0}$ as a closure ansatz for the moment hierarchy. As we shall see below, this closure is well-controlled in the small activity expansion. Upon integrating out the polarization, by rotational symmetry, we have $\rho=\rho(r)$, $\b{w}=w(r)\hat{\b{r}}$ and $\b{M}=A(r)\b{1}+B(r)\hat{\b{r}}\hat{\b{r}}$, where $r=|\b{r}|$ and $\hat{\b{r}}=\b{r}/r$.

At equilibrium ($v=0$), $w=0$, $B=0$ and all higher order odd moments vanish as well. Hence for small activity, we can estimate that to leading order $\langle\b{e}\b{e}\b{e}\rangle\sim v\rho/r$ at least, as any anisotropy must derive from $\bm\del\mathcal{U}\sim 1/r$. Similarly, the anisotropic part of $\b{M}$, $B$ must also decay at least as fast as $\sim v^2\rho/r^2$ at leading order in $v$ (it is even in $v$ as $\b{M}$ is even in $\b{e}$). We shall use these estimates below to argue their neglect.

We now wish to eliminate the fast dynamics of $\b{w}$ and $\b{M}$ to obtain a steady state solution for the density $\rho$.
We let $\tr(\b{M})=2A+B\equiv m$ and obtain an equation for $m$ as
\begin{equation}
	\partial_tm=2\mu\bm\del\cdot(\bm\del\mathcal{U}m+T\bm\del m)-\dfrac{5\mu^2\gamma|\bm\del\mathcal{U}|^2}{4K^2}(K m-2T\rho)\ .
	\label{eq:m}
\end{equation}
Balancing the order of various terms in this equation, we obtain that $m\sim\rho$ in terms of scaling. Comparing this against the estimate for the anisotropic part $B\sim v^2\rho/r^2$, we find that $B$ is negligible at large distances and small activity and hence it can be safely discarded. In addition, including third order moments, they all enter Eq.~\ref{eq:m} with an additional factor of $\sim v/r$ (from either active advection or the active torques), which in total combine to give a correction of order $\sim v^2\rho/r^2$. Just like $B$, this contribution is also negligible compared to the terms retained and justifies our closure scheme to leading order in activity and at large distances.

At large distances, the fast mode $\b{w}$ relaxes on time scales $\sim D_R^{-1}$ and can be slaved to $m$ with the result by neglecting $\partial_t\b{w}$, to leading order in the gradients,
\begin{equation}
	\b{w}=-\dfrac{v}{2D_R}\left[\bm\del m-\dfrac{3\mu\gamma}{4K}m\bm\del\mathcal{U}\right]+\mathcal{O}\left(v\dfrac{m}{r^3}\right)\ .
\end{equation}
This equation can then be used to also eliminate $\b{w}$ from the density equation, with the result
\begin{equation}
	\partial_t\rho=2\mu\bm\del\cdot\left[\left(\rho-\dfrac{3\gamma v^2}{16 K D_R}m\right)\bm\del\mathcal{U}+T\bm\del\rho+\dfrac{v^2}{4\mu D_R}\bm\del m\right]\ .
	\label{eq:rhoeff}
\end{equation}

We now seek a steady state solution of the coupled equations (\ref{eq:m}) and (\ref{eq:rhoeff}) by letting $\partial_t m=\partial_t\rho=0$.
Since we are intersted in the behavior for large pair separation, we seek to solve the equations to leading order in the gradients. Na{\"i}vely one may be tempted to write a solution of Eq. (\ref{eq:m}) as $m_{ss}=(2T/K)\rho_{ss}$, but the fact that the last term on the right hand side of Eq.~(\ref{eq:m}) is proportional to $|\bm\del\mathcal{U}|^2$ invalidates this simple approximation.
We then have the following two coupled equations for $\rho_{ss}$ and $m_{ss}$
\begin{align}
\rho_{ss}=\dfrac{K}{2T}\left\{m_{ss}-\dfrac{8K}{5\mu\gamma}\dfrac{\bm\del\cdot[\bm\del\mathcal{U}m_{ss}+T\bm\del m_{ss}]}{|\bm\del\mathcal{U}|^2}\right\}\ .\label{eq:rhom1}\\
	\left[\rho_{ss}-\dfrac{3\gamma v^2}{16KD_R}m_{ss}\right]\bm\del\mathcal{U}+ T\bm\del\rho_{ss}+\dfrac{v^2}{4\mu D_R}\bm\del m_{ss}=c\ ,\label{eq:rhom2}
\end{align}
where $c$ is an arbitrary constant. In the absence of currents at steady state, $c=0$. Substituting Eq.~\ref{eq:rhom1} into Eq.~\ref{eq:rhom2} and using $\mathcal{U}(r)=(\pi K/2)\ln(r/a)$, we obtain a homogeneous differential equation of Cauchy-Euler form,
\begin{align}
	5\pi^3\mu\gamma(8D_RK^2-3Tv^2\gamma) m_{ss}(r)&=8r\left[-5\pi^2Tv^2\gamma+2D_R(8(K\pi+2T)^2-5K\pi^2T\mu\gamma)\right]m_{ss}'(r)\nonumber\\
	&\quad+512D_RT(K\pi+3T)\ r^2m_{ss}''(r)+512D_RT^2\ r^3m_{ss}'''(r)\ ,
\end{align}
	where we have set $c=0$ and the prime denotes a derivative with respect to $r$. This equation is solved by a power law, $m_{ss}(r)=m_0(r/a)^{-\eta}$, where the undetermined exponent $\eta$ satisfies the following cubic equation,
\begin{equation}
	512 D_R T \eta^2(K\pi-T\eta)+8\pi^2 \eta[5v^2T\gamma+2D_RK(5T\mu\gamma-8K)]+5\pi^3\mu\gamma(3v^2T\gamma-8D_RK^2)=0\ .
\end{equation}
Of the three roots to this equation, we must only use the branch that connects to the equilibrium exponent $\eta=K\pi/2T$ when $v=0$. Perturbatively expanding the exponent $\eta$ around $v=0$, we have
\begin{equation}
	\eta=\dfrac{K\pi}{2T}-\dfrac{\pi v^2}{4\mu D_RT}\left(1+\mu\gamma\dfrac{3T}{4K}\right)+\mathcal{O}(v^4)\ .
	\label{eq:x}
\end{equation}
	The next order correction to $\eta$ at $\mathcal{O}(v^4)$ will also involve the neglected terms $B$ and the third moment $\langle\b{e}\b{e}\b{e}\rangle$.
Using the power law form of $m_{ss}(r)$ in Eq.~\ref{eq:rhom1} we get
\begin{equation}
	\rho_{ss}(r)\propto (r/a)^{-\eta}\ ,
\end{equation}
	as well, with the overall prefactor fixed by normalization ($\int\dd^2r\ \rho_{ss}(\b{r})=1$). Note that $\eta>2$ for the distribution to be normalizable. This immediately gives the effective potential defined as $\mathcal{U}_{\mathrm{eff}}(\b{r})=-T\ln\rho_{ss}(\b{r})$ to be
 \begin{equation}
	 \mathcal{U}_{\mathrm{eff}}(\b{r})=\eta T\ln\left|\dfrac{\b{r}}{a}\right|\ ,
 \end{equation}
	with $\eta$ given by Eq.(\ref{eq:x}), as quoted in the main text ($\eta T\equiv \pi K_{\mathrm{eff}}/2$). Correspondingly, the unbinding transition can be directly inferred from $\rho_{ss}$ by when $\eta\leq 4$ which gives the same critical activity line as quoted in the main text.


\end{document}